\documentclass[preprint,aps,prc,nofootinbib,floatfix]{revtex4}
\usepackage{epsfig,float}
\begin{document}

\title{Inclusive pion double charge exchange in light $p$-shell
nuclei}

\author{W. Fong}
\altaffiliation[Present Address: ] {Varian Medical Systems, Palo Alto, CA 
94304}
\author{J. L. Matthews}
\author{M. L. Dowell}
\altaffiliation[Present Address: ] {Optoelectronics Division, National 
Institute of Standards and Technology, Boulder, CO 80305}
\author{E.~R.~Kinney}
\altaffiliation[Present Address: ] 
{Department of Physics, University of Colorado, Boulder, CO 80309}
\author{T. Soos}
\altaffiliation[Present Address: ] {1255 Marina Point, Casselberry, FL 
32707}
\author{M. Y. Wang}
\altaffiliation[Present Address: ] {The RAND Corporation, Santa Monica, CA 90406}
\author{S. A. Wood}
\altaffiliation[Present Address: ] 
{Thomas Jefferson National Accelerator Facility, Newport News,  VA 23606}
\affiliation{Department of Physics and Laboratory for Nuclear Science \\ 
Massachusetts Institute of Technology, Cambridge MA 02139}

\author{P. A. M. Gram}
\altaffiliation[Present Address: ] 
{82-1021 Kaimalu Place, Captain Cook, HI 96704}
\affiliation{Los Alamos National Laboratory, Los Alamos NM 87545}

\author{G. A. Rebka, Jr.}
\author{D. A. Roberts}
\altaffiliation[Present Address: ] 
{Department of Radiation Oncology, University of Michigan Medical School, 
Ann Arbor, MI 
48109}
\affiliation{Department of Physics, University of Wyoming, Laramie, Wyoming
82071}

\date{\today}

\begin{abstract}
We report the results of a series of measurements of the differential
cross sections for inclusive pion
double charge exchange in $^{6,7}$Li, $^9$Be, and $^{12}$C for
positive and negative incident pions of energies 120, 180, and 240 MeV.
The data are compared with the predictions of an intranuclear cascade 
model and a model based on two sequential single charge exchange 
processes. 
\end{abstract}


\maketitle

\section{INTRODUCTION}

The study of pion double charge exchange (DCX) has been pursued with the
aim of elucidating the mechanisms of multiple pion interactions in nuclei.
Whatever the details of the mechanism, this reaction must at least result
in the charges of two nucleons being changed. Indeed, it has been found
that the simplest picture of the reaction, in which it 
proceeds by successive single charge exchanges (see Fig.\ 1), provides 
a qualitative
explanation of the systematics that are observed \cite{grametal}. 

Serious investigation of the inclusive DCX reaction began in 1980, with the
first measurement using modern equipment of the doubly differential cross
section for $^{16}$O$(\pi^+,\pi^-)$X at three angles with 240 MeV pions
performed at SIN (now PSI) \cite{mischke}. Subsequently, an extensive 
series of
measurements of inclusive double charge exchange cross sections was carried out
at the Los Alamos Meson Physics
Facility \cite{wood,yuly,kprl,kth,kinney,fongth,fong}. The $(\pi^-, 
\pi^+)$
reaction was studied in $^3$He, and the $(\pi^{\pm}, \pi^{\mp})$ reactions in
nine nuclei:  $^4$He, $^{6,7}$Li, $^9$Be, $^{12}$C, $^{16}$O, $^{40}$Ca,
$^{103}$Rh, and $^{208}$Pb. Doubly differential cross sections were measured
for each target over the outgoing pion energy range from 10 Mev up to the
kinematic limit, at three to five angles for incident pion energies in
the range from 120 to 270 MeV. Some 300 pion energy spectra have
been obtained with good statistics and absolute accuracies of 10-15\%.

The systematics of the reaction in $^{16}$O, $^{40}$Ca,
$^{103}$Rh, and $^{208}$Pb revealed 
by these experiments has been previously discussed \cite{wood}. Briefly, 
it is found
that only in 
$^{16}$O do the doubly differential cross sections much resemble the most 
likely phase space distribution (that containing a pion, two nucleons and 
a residual nucleus) or, in fact, any other reasonable distribution 
generated by assuming that the mechanism is governed by a constant 
matrix element. In the heavier nuclei the distribution of outgoing pions 
is strongly peaked toward lower energies and depleted at higher energies 
when compared to the phase space distribution.

The doubly differential cross sections for DCX in $^3$He and $^4$He
exhibited striking 
double peaked outgoing pion energy spectra at forward angles
(25$^\circ$, 50$^\circ$) that evolved into a single 
peak distribution as the angle of observation reached 130$^\circ$. This 
feature was 
most pronounced at an incident energy of 240 MeV, but persisted at
incident energies down to 150 MeV in $^4$He \cite
{kinney} and 120 MeV in $^3$He \cite{yuly}.

In this paper we report measurements of the doubly differential
cross section for inclusive DCX
in the light $p$-shell nuclei $^{6,7}$Li, $^9$Be, and $^{12}$C.  These
observations will serve to
trace the evolution of the double peaked structure into a 
nearly featureless spectrum as the complexity of the target nucleus is
increased.  A second objective of the present work was to investigate, by
comparing the $(\pi^+, \pi^-)$ reactions in $^6$Li and $^7$Li,
the effect on the
DCX cross section of a single extra neutron in the target nucleus.

Only fragmentary data for DCX in these nuclei existed prior to the present
measurement.  Gilly {\it et al.} have reported a few
measurements at 0$^\circ$ for $(\pi^+,\pi^-)$ in Li and Be \cite{gil64} 
and
$(\pi^-,\pi^+)$ in $^7$Li, $^9$Be, and $^{12}$C \cite{gil65}, at incident
energies in the 200 MeV region.  Batusov {\it et al.} measured total
DCX cross
sections for 80 MeV $\pi^+$ on Be and Al and 140 MeV $\pi^-$ on Be and
C \cite{bat66}.  Using a propane bubble chamber, Massue {\it et
al.} measured total
cross sections for $(\pi^+,\pi^-)$ in $^{12}$C at 138, 204, and 228
MeV \cite{mas71}.  Evseev {\it et al.} measured a $\pi^+$ spectrum at
30$^\circ$ for 107
MeV $\pi^-$ on $^7$Li, but with poor statistical accuracy \cite{evs81}.

Subsequent to the measurements discussed in the present paper,
forward-angle cross sections for the $(\pi^-, \pi^+)$ reaction on
$^{6,7}$Li and $^{12}$C at incident energies 0.59, 0.75, and 1.1 GeV have
been reported \cite{abramov}.

\section{EXPERIMENTAL APPARATUS AND PROCEDURE}

In the present experiment, the inclusive DCX
processes $(\pi^{\pm}, \pi^{\mp})$ have been
investigated by measuring the doubly differential cross sections, $d^2 
\sigma / d\Omega dE_\pi$, at three to five angles in the range
25$^\circ$--130$^\circ$, for incident pion energies between 120 and 
240 MeV.  At each angle, the range of outgoing pion energies from 10 MeV up to 
the kinematic limit for each reaction was covered.
Preliminary results of this experiment were reported in Ref.
\cite{fong}.

The experiment was performed with incident
$\pi^\pm$ beams from the high-energy pion channel (``P$^3$'') at the 
Los Alamos Meson Physics Facility. 
The outgoing pions were detected using a 180$^\circ$ vertical bend, double
focusing magnetic spectrometer, with an effective solid angle of about 16
msr and a momentum bite of about 8\%, as shown in Fig.\ 2. The experimental
apparatus and procedure have been discussed in detail by Wood {\em et
al.}\ \cite{wood}, Yuly {\it et al.} \cite{yuly}, and Kinney {\it et al.} 
\cite{kinney} and will be treated
only briefly here. 

The detector system had five components: a wire chamber (WC0) placed in
the mid-plane of the spectrometer, two wire chambers (WC1 and WC2) near
the focal plane that were used to reconstruct trajectories, a 1.6 mm
scintillator (S1) behind WC1 and WC2 that provided an accurate time 
reference for the trigger
as well as pulse height information, and a fluorocarbon (FC-88) \v
Cerenkov detector that distinguished electrons from pions.

The trigger was a fourfold coincidence among the three wire chambers and
the scintillator.  Inclusion of WC0
guaranteed that a particle had passed through the spectrometer. The timing
information available from this chamber was also used to distinguish pions
from very slow protons.

All of the targets ($^6$Li, $^7$Li, $^9$Be, $^{12}$C) in this experiment 
were solid rectangular slabs.  In
addition, a
polyethylene (CH$_2$) target was employed to simultaneously normalize both 
the beam
flux and spectrometer acceptance by observing elastic $\pi p$
scattering.  The targets were mounted in aluminum frames on a
ladder that could be raised or lowered (to place the target of interest in
the incident beam) and rotated about the vertical axis (to orient the 
plane of the 
target at an appropriate angle with respect to the beam).

Before reaching the target, the pion beam passed through an ionization
chamber that was used to determine the relative flux. Since this device was
sensitive to all charged particles in the beam, it was necessary to
normalize the flux measurement each time the beam transport elements were
adjusted.  Downstream of the target the pion beam impinged on a  
CH$_2$ slab from which pions scattered at 90$^\circ$ on each side of the beam 
were detected by scintillator telescopes, in order to 
monitor the position of the incident beam and to provide a cross-check 
of the ionization chamber.

To measure the doubly differential DCX cross sections each target was
exposed to an incident pion beam and events were recorded at a series of
spectrometer momentum settings until the measured statistical 
uncertainties in the
number of detected pions were 
approximately 5\%. Data were collected at 10 MeV intervals in outgoing
pion energy. In addition, background observations
were made at a selected set of spectrometer settings using an empty target 
frame.  Each time the
incident pion charge or energy was changed, a series of CH$_2$
normalizations was performed.

\section{DATA ANALYSIS}

The goal of this experiment was to measure a doubly differential cross
section for each reaction, incident pion energy, scattering angle, and
outgoing pion energy. The doubly differential cross section is
related to observable quantities as follows:
\begin{equation}  \label{eqn:xsect}
\frac{d^2 \sigma}{d \Omega \, d E_\pi}= \frac{N_{{\rm det}} \, \epsilon_c}{
N_{{\rm inc}} \, x \, \rho \, f_d \, f_l \Delta \Omega \, \Delta E_\pi }
\end{equation}
where $N_{{\rm det}}$ is the number of pions detected, $N_{{\rm inc}}$ is
the number of incident pions, $x$ is the effective thickness of
the target, $\rho$ is the target density, $\Delta \Omega$ is the
effective solid angular acceptance, $\Delta E_\pi$ is the range of outgoing
pion energy, $\epsilon_c$ is the correction of the spectrometer acceptance
due
to multiple scattering and energy loss, $f_d$ is the correction due to pion
decay, and $f_l$ is the dead-time correction.

The analysis procedure is described in detail in Ref.~\cite{fongth} and will
only be summarized here. (See also Refs.~\cite{yuly,kinney}.) 

\subsection{Wire chamber calibration and phase space definition}

Calibration constants relating time differences of the signals from
the wire
chambers to positions on the focal plane were established by placing a
collimated $^{55}$Fe
source at precisely measured positions in front of the chambers. In the 
data analysis, the position information from WC1 and WC2 for each event 
was used to reconstruct
the
particle trajectory back to the focal plane of the spectrometer.
Reconstructed trajectories were tested for conformity with the distribution
in phase space of particles that could have been transmitted by the
spectrometer from the
target
to the focal plane.

\subsection{Particle Identification}

Positive pions and protons were generally distinguished on an
event-by-event basis
by means of their different pulse heights in the scintillator, or, in some 
cases, by their time-of-flight between WC0 and S1.

Since pions and electrons\footnote{
Electrons or positrons, depending on the charge setting of the
spectrometer.
The word ``electron'' will be used generically throughout this discussion.}
both emit \v {C}erenkov radiation in FC-88 at spectrometer settings 
greater than 180
MeV/c
($T_\pi =88$ MeV), it was not possible to use the \v {C}erenkov detector to
distinguish electrons from pions on an event-by-event basis. Instead, a
statistical method was used, which is described in detail in Ref.~\cite{fongth}
and is very similar to that described in Ref.~\cite{yuly}.

\subsection{Acceptance and Dispersion}

The momentum acceptance, $\Delta p/p$, and dispersion were determined by
changing the spectrometer magnetic field to move a $\pi p$ elastic
scattering peak across the spectrometer focal
plane.  The
relative acceptance as a function of focal plane position obtained by
this method was used to correct $N_{\rm det} $. The effective total 
momentum
acceptance was determined to be about 8\% by integration across the entire
focal plane. 

\subsection{Normalization}

The absolute number of incident pions, $N_{{\rm inc}}$ in Eq.~(\ref
{eqn:xsect}), was obtained by comparison of $\pi p$ elastic scattering
measurements with known cross sections determined
from the energy-dependent phase shift program
SCATPI~\cite{walter}.
The effective total solid angle, $\Delta \Omega$, was included in this
calibration. This procedure
was repeated at several scattering angles for each setting of the beam
transport system in order to improve the accuracy of the normalization, as
well as to check for the presence of angle-dependent systematic errors.

\subsection{Background Subtraction}

A linear interpolation between the spectrometer momentum settings at which
empty-frame (background) measurements were made was used to obtain the
background contribution which was then subtracted from the raw DCX
spectra.
The largest backgrounds occurred at the lowest measured outgoing pion 
energy of 10
MeV and were
about 25\% of the foreground.

\subsection{Corrections}

Corrections were made to account for other effects that would change the
shape and magnitude of the measured cross sections. The number of pions
detected was reduced by decay as they traveled from the target to the
detectors. Approximately 80\% of the pions survived at 200 MeV, while at 10
MeV only 30\% survived. Some of the decay muons traversed the spectrometer,
while others did not. Moreover, it was possible for pions emerging from 
the target that would not
have traversed the spectrometer to decay into muons that could do so. Since
the detector system could not separate the muons from pions, these effects
were accounted for by Monte Carlo methods which produced the factor $f_d$
in Eq.\ (\ref{eqn:xsect}). This procedure is described in detail in Refs.\
\cite{wood,wth}.  The fraction of muon contamination ranges from 1\% at 10 
MeV to 20\% at 200 MeV outgoing pion kinetic energy.  

Energy loss and multiple scattering in the target and in WC0
changed the effective acceptance of the spectrometer; these were corrected
for by the factor $\epsilon_c$, which was determined by simulation to be
different from unity only for outgoing pion energies less than 30 MeV. The
maximum value of $\epsilon_c$ was about 1.3.

\subsection{ Integrated cross sections}

Integration of the doubly differential cross sections over outgoing pion
energy, using the trapezoid rule, yielded an angular distribution.  The angular
distributions were fitted with sums of Legendre polynomials, which were
integrated to determine a quantity which we term the ``total reaction cross
section.'' The uncertainties in these procedures are discussed in Ref.  
\cite{fongth} and were less than, or comparable to, the systematic
uncertainties discussed in the next section.

\subsection{Systematic uncertainties}

The uncertainty in the normalization of the cross section contains 
contributions
from uncertainties in the thicknesses of the targets (2--3\%) and in the phase
shift prediction of the free $\pi p$ cross section (2\%) \cite{walter}.  
The
uncertainty in the determination of a particle's position in the spectrometer
focal plane and of the spectrometer acceptance function are each estimated to be
1\%. For the measurements at 180 MeV and 240 MeV there is an additional 4\% 
uncertainty in a correction for trigger inefficiency.  The uncertainties due to 
the determination of electronics deadtime,
 spectrometer dispersion, and wire chamber efficiency were insignificant
compared to those listed above.  In addition, there is an uncertainty due
to slowly varying changes in the response of the ionization chamber used
to monitor the incident beam flux. These variations were monitored by
comparison with the primary proton beam monitor and the downstream
scattering monitor, and were observed to occur on roughly the same time
scale as changes of spectrometer angle.  The resulting
uncertainty is thus treated as an angle-dependent uncertainty.

There are also uncertainties that depend on the outgoing pion energy.
These include contributions from the electron-pion separation procedure,
which were between 2\% and 5\%, from
estimates of the corrections for energy loss and multiple scattering in
the target and in the mid-spectrometer wire chamber, and from effects of
pion decay and muon contamination, which contributed an additional 5\%.

Table I lists the systematic uncertainties for each
incident beam.

\subsection{Pion-induced pion production}

For incident pion energies above about 170 MeV, detection of an outgoing pion of
charge opposite to that of the incident pion no longer provides a unique signature
of DCX, as the processes $n(\pi^+,\pi^-)\pi^+p$ or $p(\pi^-,\pi^+)\pi^-n$ can
occur on one of the target nucleons.  Although the contribution of pion-induced
pion production (PIPP) is negligible (see Ref.~\cite{wth}) for incident energy 180
MeV, this effect must be taken into account at 240 MeV.

To estimate the contribution of PIPP, we follow the procedure used in
Ref.~\cite{kinney}.  Assuming that PIPP in a nucleus occurs as a
quasi-free process on a single nucleon, the effects of absorption and
other competing $\pi$-nucleus reactions and of the nuclear medium are
taken into account by multiplying the free nucleon PIPP cross section by
an ``effective'' number of nucleons, $N_{\rm eff}$.  An extensive study of
$\pi$-nucleus reactions was reported by Ashery {\it et al.}\cite{ashery}, 
from which values of $N_{\rm eff}$ were extracted.  
These authors found that $N_{\rm eff}$ for positive (negative) incident pions 
exhibited a universal, energy-independent power-law dependence on $Z$ 
($N$).  Moreover, the incident energy dependence of $N_{\rm eff}$ was 
found to be very similar for all target nuclei.  These two results allow 
us to obtain, by interpolation, values of $N_{\rm eff}$ for 
$^{6,7}$Li, $^9$Be, $^{12}$C, and $^{16}$O for incident $\pi^{\pm}$ energy 
240 MeV.

Precise measurements of the $\pi^- p \rightarrow \pi^- \pi^+ n$ cross 
section have been reported 
by Bjork {\it et al.}\cite{bjork}.  The observed\cite{piasetzky} equality 
of the $\pi^+ d \rightarrow \pi^+ \pi^- pp$ and $\pi^- d \rightarrow 
\pi^- \pi^+ nn$ cross sections, and of the $\pi^- d \rightarrow \pi^- 
\pi^+ 
nn$ and $\pi^- p \rightarrow \pi^- \pi^+ n$ cross sections, justifies 
using the results of Bjork {\it et al.}\cite{bjork} for both positive and 
negative incident pions.  This procedure yields the values of PIPP cross 
sections given in Table II.

Although extensive studies of the differential cross sections for PIPP in
nuclei have been carried out, {\it e.g.}, most recently by the CHAOS
collaboration \cite{chaos}, few total cross sections have been quoted.  
Rahav {\it et al.} \cite{rahav} observed the $^{12}$C$(\pi^-, \pi^+ 
\pi^-)$
reaction at 292 MeV, and Grion {\it et al.} \cite{grion} observed the
$^{16}$O$(\pi^+, \pi^+ \pi^-)$ reaction at 280 MeV.  Unfortunately, these
results are inconsistent when rescaled using a reasonable energy and
$A$-dependence (see Ref.~\cite{wood}).  Rahav {\it et al.} speculate that
the discrepancy may arise from the methods used to extrapolate the data to
obtain a total reaction cross section.  They point out that the fraction
of phase space covered in their work is 43 times larger than that covered
in the experiment of Grion {\it et al.}.  If we use the energy dependence
of the free nucleon PIPP cross section \cite{bjork} to rescale Rahav {\it
et al.}'s result for $^{12}$C to 240 MeV, we obtain $206 \pm 26 \mu$b, in 
fair
agreement with the value obtained using $N_{\rm eff}$ given in Table II.

We note that PIPP can only affect the low energy portion of the outgoing
pion energy spectra.  The estimated contributions will be shown when the 
results
are presented in the next section.

\section{RESULTS}

\subsection{Doubly differential cross sections}

A sample of the results of the present
investigation will be displayed in this section\footnote{The complete set of data may be found in 
Ref.~\cite{fongth}.} and some qualitative and semi-quantitative
aspects of the measured
cross sections will be discussed.  The data will be compared with the predictions of two
theoretical models in the next section.  

Doubly differential cross sections for the $^6$Li$(\pi^+, \pi^-)$ process are  
shown in 
Figs.~3, 4, and 5 for incident energies 120, 180, and 240 MeV, 
respectively, and angles of observation 
between 25$^\circ$ and 130$^\circ$.    The $^6$Li$(\pi^-, \pi^+)$ cross sections 
(see Ref.~\cite{fongth}) are essentially identical in shape and magnitude and are 
not 
shown.  The solid curves represent the distribution of events in four-body 
(outgoing pion, two nucleons, and residual nucleus) phase space 
and are normalized in each case to the area under the experimental spectrum.
The phase space curves are seen to provide a good representation of the data at 
incident pion energy 
120 MeV and a poorer representation at the higher incident energies.
The dashed curves in Fig.~5 indicate the estimated contribution of 
PIPP, assuming a phase-space distribution (as was found by Rahav {\it et 
al.} \cite{rahav}).  The angular dependence of the PIPP cross section was 
obtained from the measurements of Yuly {\it et al.} \cite{yuly} and 
the normalization determined by the total cross section given in Table 
III.

Figure 6 shows the doubly differential cross sections for $^7$Li$(\pi^+,
\pi^-)$ (solid symbols) and $^7$Li$(\pi^-, \pi^+)$ (open symbols) at 240
MeV incident energy.  The dotted and dot-dashed curves indicate the
estimated contributions of PIPP to the $(\pi^+, \pi^-)$ and $(\pi^-,
\pi^+)$ cross sections, respectively.  Comparing Figs.~5 and 6, it is seen
that the shapes of the energy spectra are very similar at 240 MeV for the
two Li isotopes.  This similarity persists at the lower incident energies
\cite{fongth}.  Although the $^7$Li$(\pi^+, \pi^-)$ and $(\pi^-, \pi^+)$
cross sections are essentially identical in shape, the former is seen to
be approximately a factor of three larger than the latter.  This
difference in magnitude will be discussed later in this section.  It is 
also seen that the PIPP contribution is nearly identical for both 
reactions, and thus is relatively larger for $^7$Li$(\pi^-, \pi^+)$.

Figure 7 shows the doubly differential cross sections for $^9$Be$(\pi^+,
\pi^-)$ (solid symbols) and $^9$Be$(\pi^-, \pi^+)$ (open symbols) at 180
MeV incident energy.  The $(\pi^+, \pi^-)$ cross section is seen to be
approximately a factor of three larger than that for $(\pi^-, \pi^+)$.

As was stated earlier, one motivation for the study of inclusive DCX in light
$p$-shell nuclei was to trace the $A$-dependence of the cross section -- in
particular the forward-angle cross section -- from the double-peaked 
structure
seen in $^{3,4}$He to the single peak seen in $^{16}$O and heavier nuclei.  
Figure 8 shows the doubly differential cross sections at 25$^\circ$ for 
240 MeV
(Fig.~8(a)) and 180 MeV (Fig.~8(b)) positive pions incident on $^4$He, 
$^6$Li,
$^7$Li, $^9$Be, $^{12}$C, and $^{16}$O.  Indeed, one observes a smooth
progression in the shape of the spectra as $A$ increases.\footnote{A very similar
$A$-dependence is observed in the $(\pi^-, \pi^+)$ cross sections.} At 240 MeV,
the double-peaked structure is prominent in $^4$He, apparent in the Li 
isotopes, disappearing for
$^9$Be, and is absent in $^{12}$C and $^{16}$O.  At 180 MeV the 
double-peaked structure is only seen in the $^4$He spectrum, although the 
data for $^{6,7}$Li and $^9$B exhibit strength
at high energy that does not appear in the $^{12}$C and $^{16}$O spectra.  
We note that in $^4$He
the double peaks are still visible at 150 MeV \cite{kth,kinney}, and in $^3$He 
at all energies down to 120 MeV \cite{yuly}.

It is seen that the relative size of the PIPP contribution, shown as the 
dotted curves in Fig.~8(a), 
decreases with increasing $A$.

The origin of the double-peaked structure has been discussed previously
(see, {\it e.g.}, Refs.~\cite{yuly,kinney}), and has been argued to be
consistent with a sequential single charge exchange mechanism.  In the
$\Delta$ resonance region the $\pi$-nucleon cross section is strongly
forward- and backward-peaked.  Two forward scatterings or two backward
scatterings, which are favored over scattering at intermediate angles, can
yield an outgoing DCX pion at forward angles:  in the former case the pion
emerges with high energy; in the latter case it emerges with lower energy.  
A theoretical model based on these ideas was developed by Kinney and
successfully applied to DCX data on $^4$He \cite{kth,kinney} and,
subsequently, $^3$He \cite{yuly}.  Given the prominent double-peaked
structure in the $^6$Li cross section, we have extended the model to this
nucleus, and will show the results in the next section.

Alqadi and Gibbs \cite{ag} have presented calculations for DCX in
$^4$He which they compare with the data given in Ref.~\cite{kth}.  These authors
found that including PIPP in their intranuclear cascade model was essential to
produce reasonable agreement with the data at 240 MeV, and suggest that this 
mechanism
is largely responsible for the low-energy peak in the pion energy spectra.  
The contribution of PIPP for $A \ge 4$, obtained as discussed 
previously, is shown by the dashed curves in Fig.~8(a).

\subsection{Angular distributions}

As described in the previous section, the doubly differential cross
sections were integrated to yield angular distributions (see
Ref.~\cite{fongth}).  A typical set of results, those for $^6$Li, is shown
in Fig.~9.  For this nucleus and the other nuclei investigated, the 
angular
distribution is seen to be forward peaked at 240 MeV and to become more
isotropic as the incident energy decreases.  No significant difference is
seen in the angular dependences of the $(\pi^+,\pi^-)$ and $(\pi^-,\pi^+)$
cross sections in those nuclei for which both were measured.  Although the
angular distributions are relatively featureless, their shape does exhibit
a consistent $A$-dependence, becoming less forward-peaked as $A$
increases.  As a measure of the forward-peaking, the ratios of the cross
sections measured at $25^\circ$ and $130^\circ$ for the
$(\pi^\pm,\pi^\mp)$ reactions at 240 MeV for $3 \leq A \leq 208$ were
calculated and are displayed in Table III.  The observed trend is
consistent with the increased probability of absorption of forward-going
pions as $A$ increases;  the DCX pion has a greater chance of escaping
from the ``rear'' of a heavy nucleus.  The unusually small ratio seen for
$^{208}$Pb$(\pi^-,\pi^+)$ may be attributable to the large excess of
neutrons, with which the incoming $\pi^-$ preferentially interacts, in
this nucleus.

\subsection{Total reaction cross sections}

Total 
DCX reaction cross sections were determined as described in the previous 
section; the results for the light $p$-shell nuclei are 
given 
in Table IV. The quoted errors include statistics, systematics, and the 
uncertainties arising from the integration procedures.  The numbers in 
parentheses for 240 MeV incident pions are the 
cross sections after the subtraction of the estimated contributions of 
PIPP.

For a self-conjugate nucleus, the $(\pi^+,\pi^-)$ and $(\pi^-,\pi^+)$ cross 
sections should be equal, apart from differences attributable to the different 
binding energies of the final state nuclei, which would affect primarily the 
high-energy 
end of the outgoing pion energy spectrum.  The total cross sections for DCX in 
$^{16}$O 
and $^{40}$Ca were found to be equal within the experimental uncertainties 
\cite{wood}, although the 
$(\pi^-,\pi^+)$ cross section slightly exceeded the $(\pi^+,\pi^-)$ cross 
section for $^{40}$Ca.  Table IV shows the two cross sections for $^6$Li 
to be 
essentially equal within errors.

\subsection{$A$-dependence of cross sections}

Starting from the notion that double charge exchange proceeds by
sequential single charge exchange scatterings competing with more probable
reactions that either interrupt the sequence or scatter pions out of the
nucleus before it starts, a ``scaling rule''was found which organizes the
$A$-dependence of the total cross section for both the $(\pi^+,\pi^-)$ and
$(\pi^-,\pi^+)$ reactions \cite{grametal}.  According to this rule the
total cross section for DCX should vary as \begin{equation} \sigma \sim
A^{2/3}Q(Q-1)/(A-Q)(A-1), \end{equation} where $Q$ is the number of
nucleons ($N$ or $Z$) on which DCX occurs for a given incident pion 
charge (positive or negative); the
factor $A^{2/3}$ is proportional to the projected area of the nucleus.
Total DCX reaction cross sections in nuclei ranging in mass from $^6$Li to
$^{208}$Pb are found substantially to obey this rule; only the isotopes of
He violate it \cite{grametal}.  Given that the scaling rule is
based on the classical transport of pions through nuclear matter, it is
perhaps surprising that the agreement is so good, especially for the light
nuclei.  The validity of the rule has recently been discussed by Buss {\it 
et al.} \cite{buss}.

Let us examine the predictions of the scaling rule in more detail for the 
light
$p$-shell nuclei. By comparing the $(\pi^{\pm}, \pi^{\mp})$ cross sections
in $^6$Li and $^7$Li, one can examine the effect of adding one neutron.  
For $(\pi^+,\pi^-)$ one is adding a target nucleon; for $(\pi^-,\pi^+)$
one is adding a source of
competing reactions and the number of target nucleons is unchanged. 
Eq. (2) predicts the ratios $R(\pi^+,\pi^-) \equiv
\sigma(\pi^+,\pi^-)_{^7{\rm Li}} / \sigma(\pi^+,\pi^-)_{^6{\rm Li}} =
1.85$ and $R(\pi^-,\pi^+) \equiv \sigma(\pi^-,\pi^+)_{^7{\rm Li}} /
\sigma(\pi^-,\pi^+)_{^6{\rm Li}} = 0.69$.  Note that in the absence of
competing reactions $R(\pi^-,\pi^+)$ would be equal to unity.  The
observed ratios are given in Table V.  The agreement between experiment
and prediction for $R(\pi^-,\pi^+)$ is excellent at both 180 and 240 MeV;
the observed values of $R(\pi^+,\pi^-)$ slightly exceed those predicted.

We can further examine the rule by comparing the $(\pi^+,\pi^-)$ 
and $(\pi^-,\pi^+)$ 
reactions in the $N \neq Z$ nuclei.  In $^7$Li and $^9$Be, the extra neutron 
should 
enhance the $(\pi^+,\pi^-)$ 
reaction by providing more targets for the charge-exchange reactions and 
suppress the $(\pi^-,\pi^+)$ reaction by providing more targets for the 
competing reactions.  
Eq. (2) predicts ratios of 
$(\pi^+,\pi^-)$ to $(\pi^-,\pi^+)$ of 2.67 in $^7$Li and 2.08 in $^9$Be. 
The experimental ratios of the total cross sections, given in Table VI, 
are seen to 
agree well with these predictions at 240 MeV but to exceed them at 180 MeV.

To avoid additional uncertainties and assumptions, the ratios in Tables V 
and VI were obtained from the measured cross sections without the subtraction of 
PIPP.  It was found that performing this subtraction does not alter the 
ratios outside their uncertainties and thus does not alter the conclusions 
drawn.

\section{Comparison of Data with Theoretical Models}

\subsection{Intranuclear Cascade Calculation}

In the intranuclear cascade calculation of Oset and collaborators
\cite{oset,salcedo,vicente,vo}, the probabilities for pion quasi-elastic
scattering and absorption are computed using a microscopic model in 
which the
pion is propagated classically through the nucleus.  
Quasi-elastic single charge exchange as well as non-charge exchange 
interactions are
allowed, and thus this model can be employed to generate DCX cross
sections based on the sequential single charge exchange mechanism.  The 
calculation is performed in
infinite nuclear matter, invoking the local density approximation to
evaluate the reaction probabilities.  Further details of the calculation
may be found in Ref.~\cite{fongth}.
Results of this calculation have previously been compared with DCX data on
$^{16}$O, $^{40}$Ca, $^{103}$Rh, and $^{208}$Pb \cite{wood,vicente,vo}.  
Ref.~\cite{vo} also contains some comparisons with results of the present
experiment for $^9$Be and $^{12}$C.  

The predictions of this model are compared with the present 
data for $^6$Li$(\pi^+,\pi^-)$ and $^7$Li$(\pi^+,\pi^-)$ at 240 MeV  in 
Figs.~10 and 11, respectively.  A persistent feature of the 
calculations is a low-energy peak that is not 
seen in the data.  A hint of the double-peaked structure seen in the 
forward-angle spectra does appear in the calculation, but the high energy peak 
is 
at too high an energy and of insufficient amplitude to match that observed. 
Comparison of this 
calculation with data on $^9$Be shows qualitatively similar results -- see 
Fig.\ 6 of Ref.~\cite{vo}.  If one attributes the measured strength in the 
low-energy region of the forward angle spectra to PIPP, as suggested by the work 
of Alqadi and Gibbs \cite{ag}, then the disagreement between the measured and 
calculated doubly-differential cross sections is even more striking.

Given the disparities in the measured and calculated doubly differential
cross sections, comparison of the integrated cross sections may not be
very meaningful.  However, it can be seen by inspection of the energy
spectra that the magnitude of the cross section is better reproduced by
the theory for $^7$Li than for $^6$Li.  Fig.\ 12 shows the theory and
measurement for $^7$Li$(\pi^-,\pi^+)$ at 240 MeV.  The comparison of
theory and experiment is qualitatively similar to that for
$^7$Li$(\pi^+,\pi^-)$.  Calculations have also been performed
\cite{fongth} at 180 MeV and 120 MeV.  Although the calculation reproduces
the approximate isotropy of the cross sections at these lower energies,
quantitative comparisons are on the whole no more successful than at the
higher energy. But perhaps it is not surprising that a calculation
designed for heavy nuclei cannot predict the details of the DCX process in
these very light nuclei. Vicente-Vacas and Oset \cite{vo} comment that a
major weakness of their calculation is the fact that the model
\cite{salcedo} predicts narrower quasi-elastic scattering peaks than are
observed, and that this might have a bearing on the discrepancy seen
between theory and measurement for light nuclei.

\subsection{ Non-static Sequential Single Charge Exchange Model}

We next compare our data with the predictions of a model developed by 
Kinney \cite{kth} based on the sequential single charge exchange 
mechanism.  
This treatment is an extension of some of the work of van Loon \cite{vl} 
and 
relies on the formalism developed by Thies \cite{thies} to describe 
inelastic, 
multistep reactions with quantum-mechanical transport theory.  

The essential features of the calculation are discussed in
Ref. \cite{kinney}; an extended description is contained in Ref.
\cite{kth}.  
In this model the incident pion interacts sequentially with two
like-charge nucleons only, and thus
only the leading, or ``double scattering'', term in the transition
matrix is used, {\it i.e.},
\begin{equation}
T=\sum_{i=1}^A\sum_{j\neq i}^At_iG_0t_j,
\end{equation}
where the $t_i$ are the in-medium transition operators ($t$-matrices) for
scattering from
the $i$th nucleon, and $G_0$ is the in-medium pion propagator. 
What distinguishes this calculation from simpler folding models \cite{wth,vl} 
is the non-static treatment of
the $\pi N$ interaction, which is important because in the energy region of the 
$\Delta$ resonance, the interaction between
the pion and the nucleon varies strongly with pion-nucleon relative
energy. Also,
the binding of the nucleons is taken into account via an approximate prescription.

The predictions of this model have been compared with DCX data on $^3$He
\cite{yuly} and $^4$He \cite{kinney}.  Here we extend the calculations to
$^6$Li, and accordingly must modify the initial state wave function to take
into account the presence of the $p$-shell proton and neutron.  We 
describe the
$^6$Li nucleus by a product of $(1s)$ and $(1p)$ harmonic oscillator wave
functions:  $(1s)^2(1p)$.  The DCX amplitude is summed over the three possible
sequences of first and second interactions:  $(ss)$, $(sp)$, and $(ps)$.  The $(ss)$ 
sequence is seen to contribute about 25\% and the combined $(sp)$ and $(ps)$ 
sequences about 75\% to the total amplitude \cite{fongth}.

In Fig.\ 13 we compare the results of this calculation with the measured 
doubly
differential cross section at five angles for incident pion energy 240 
MeV, with two assumptions for the value of the average nuclear potentials 
$U_1$ and $U_2$ at the first and second scattering, 
respectively. The
solid curves (lower branches) represent the calculation for $U_1 = U_2 = 
-30$ MeV; the dashed
curves (lower branches) that for $U_1 = U_2 = 0$.  The sequential 
scattering model does indeed 
predict a
high energy peak in the forward angle cross sections that is reasonably close,
both in magnitude and in position in energy, to that seen in the data; the 
evolution
of this peak to larger angles is also quite well reproduced. 

Whereas for $^3$He and $^4$He this calculation predicted a double-peaked
spectrum at forward angles that agreed fairly well with the data,
the inclusion of a $p$-shell nucleon seems to greatly diminish the 
low-energy
peak, compared with the data.  The question arises as to
whether the primary origin of the observed low-energy peaks at 240 MeV in 
the light $p$-shell
DCX data is in fact PIPP, as was suggested by Alqadi and Gibbs \cite {ag}
for DCX in $^4$He.  The upper branches of the solid and dashed curves were
obtained by adding the estimated PIPP cross sections from Fig.~5 (which
assume a phase-space distribution) to the calculated DCX cross sections.  
The agreement with the data is seen to be
surprisingly good, and may be fortuitous, given the simplicity of the 
assumptions made for both DCX and PIPP.

The DCX calculations in Fig.~13 show very little dependence on the nuclear
potential.  However, when the calculation is compared to the data at
25$^\circ$ for incident energies 120, 180, and 240 MeV (Fig.\ 14), we see
that the choice of the potential has a significant effect. Whereas the
calculation for $U_1 = U_2 = 0$ provides a fairly good representation of
the magnitude of the cross section at all three energies, that for $U_1 =
U_2 = -30$ MeV produces a prediction that is approximately three times too
large at 180 MeV and five times too large at 120 MeV.  The same phenomenon
was seen in the comparison of this model with data on $^3$He and $^4$He:
only by setting both potentials to zero could the magnitude of the
predicted cross section be brought close to the data at 120 and 180 MeV.

\section{Summary and Conclusions}

A systematic study of the inclusive DCX process in light $p$-shell nuclei
has been performed for incident pion energies between 120 MeV and 240 MeV.  
The relative magnitudes of the cross sections were found to be roughly
consistent with a ``scaling rule''\cite{grametal} derived by assuming DCX 
to
proceed via two sequential single charge exchange (SSCX) reactions in
competition with other processes.

The data are compared with the predictions of two theoretical models:  
an intranuclear cascade model of Oset and collaborators
\cite{oset,salcedo,vicente,vo} and a model developed by 
Kinney \cite{kth} based on the SSCX mechanism.  Although one would expect 
the model of Oset, which
allows multiple interactions, to provide a better reproduction of the
data, this does not appear to be the case.  In particular, the strength at
low outgoing pion energies, which one would expect to arise from multiple
interactions, is calculated to be much larger than that seen in the data.  
Either the effect of these interactions has been overestimated, or one or
more of the approximations in this calculation has produced a distortion 
of the spectrum.

The calculation of Kinney \cite{kth} provides an overall better
representation of the data for $^6$Li, particularly if PIPP is included. 
Although a
double-peaked structure in the forward angle spectra is expected if SSCX
is the dominant mechanism, the structure seen for $^6$Li and $^7$Li at 240
MeV probably contains a contribution from PIPP.  We note that the data at 
180 MeV
for these nuclei do not exhibit a low-energy peak.

However, PIPP cannot provide the complete explanation, since for $^3$He 
and $^4$He the
low-energy peak persists to energies well below the PIPP threshold.  We suggest
that both SSCX and PIPP mechanisms are at work to produce pions of 
charge opposite to 
that of the incident beam.  The double peak signature
of SSCX appears clearly in $^3$He and $^4$He owing to the simplicity of these
nuclei.  As $A$ increases, the presence of additional nucleons affords the
possibility of more than two interactions, and ``washes out'' this structure.

In conclusion, the work reported in this paper completes a systematic 
study of the $A$-dependence of the inclusive double charge exchange 
process at intermediate energies.  Results for lighter \cite{yuly,kinney} 
and heavier \cite{wood} nuclei have been published previously.  Although 
many qualitative features of the inclusive DCX process are now fairly 
well understood, a quantitative theoretical description of the 
measured cross sections has yet to be achieved.

\section{ACKNOWLEDGMENTS}

This work was supported in part by funds provided by
the U.S. Department of Energy.
We acknowledge the contributions of D. Cors, S. H\o ibr\aa ten, P. Mansky,
and C. Schermer to the data-taking phase of this experiment.  We 
wish to thank
E. Oset for making his code available to us and for guidance in its use. 
We are grateful to M. Yuly 
for insightful discussions of the data analysis procedures and for a 
careful reading of the final manuscript.  T. Akdogan provided 
valuable assistance in the preparation of the figures.

\pagebreak
\clearpage



\begin{table}[tbp]
\caption{Systematic uncertainties in the experiment.}
\label{table:TableI}
\begin{ruledtabular}
\begin{tabular}{c|ccc}
\multicolumn{4}{c}{Systematic uncertainties (\%)} \\ \hline
Incident Beam & Normalization & Angle-dependent$^a$ & Overall$^b$ \\ 
\hline
120 MeV $\pi^+$ & 4.8 & 1.3 & 5.0 \\
120 MeV $\pi^-$ & 5.1 & 5.3 & 7.4 \\
180 MeV $\pi^+$ & 6.5 & 5.1 & 8.3 \\
180 MeV $\pi^-$ & 6.4 & 1.0 & 6.4 \\
240 MeV $\pi^+$ & 6.2 & 1.9 & 6.5 \\
240 MeV $\pi^-$ & 6.8 & 1.6 & 7.0 \\
\end{tabular}
\end{ruledtabular}
\footnotetext{$^a$Angle-dependent uncertainty due to variation in
ionization chamber response (see text).\\$^b$Includes angle-dependent
and normalization uncertainties only. The uncertainties which
depend on the energy of observation are included along with the statistical
uncertainties in the plotted error bars.}
\end{table}



\begin{table}[tbp]
\caption{Estimated cross sections ($\mu$b) for pion-induced pion 
production by 240 MeV incident pions.  The result for $^4$He is taken 
from Ref.~\protect\cite{kinney}.  The quoted uncertainties reflect only 
the uncertainty in the $\pi^- p \rightarrow \pi^+ \pi^- n$ cross section, 
except in the case of $^7$Li$(\pi+, \pi^-)$ where there is an additional 
uncertainty arising from the determination of $N_{\rm eff}$.} 
\label{table:TableII} \begin{ruledtabular} \begin{tabular}{c|cc}
Target nucleus & $(\pi^+, \, \pi^-)$ & $(\pi^-, \, \pi^+)$ \\ \hline
$^4$He & $128 \pm 8$  & \\
$^6$Li & $190 \pm 11$ & $190 \pm 11$ \\
$^7$Li & $252 \pm 27$ & $190 \pm 11$ \\ 
$^9$Be & $245 \pm 15$ & $219 \pm 13$ \\
$^{12}$C & $268 \pm 16$ & \\
$^{16}$O & $309 \pm 19$ & \\
\end{tabular}
\end{ruledtabular}
\end{table}



\begin{table}[tbp]
\caption{Ratio of the differential cross section at 25$^\circ$ to that at 
130$^\circ$, for 240 MeV incident pions.}
\label{table:TableIII}
\begin{ruledtabular}
\begin{tabular}{c|cc}
Target nucleus & $(\pi^+, \, \pi^-)$ & $(\pi^-, \, \pi^+)$ \\ \hline
$^3$He & & $2.61 \pm 0.19$ \\
$^4$He & $2.92 \pm 0.14$ & $2.67 \pm 0.30$ \\
$^6$Li & $1.82 \pm 0.07$ & $1.84 \pm 0.07$ \\
$^7$Li & $1.74 \pm 0.06$ & $1.66 \pm 0.06$ \\ 
$^9$Be & $1.51 \pm 0.02$ & $1.57 \pm 0.03$ \\
$^{12}$C & $1.27 \pm 0.02$ & \\
$^{16}$O & $1.18 \pm 0.04$ & $1.33 \pm 0.05$ \\
$^{40}$Ca & $0.99 \pm 0.03$ & $1.01 \pm 0.03$ \\ 
$^{103}$Rh & $0.78 \pm 0.02$ & $0.77 \pm 0.02$ \\
$^{208}$Pb & $0.82 \pm 0.02$ & $0.54 \pm 0.01$
\end{tabular}
\end{ruledtabular}
\end{table}

\pagebreak



\begin{table}[tbp]
\caption{Total DCX reaction cross sections for $^6$Li, $^7$Li, $^9$Be, and 
$^{12}$C.}
\label{table:TableIV}
\begin{ruledtabular}
\begin{tabular}{c|cccc}
\multicolumn{5}{c}{Reaction cross sections (mb)} \\ \hline 
Incident Beam & $^6$Li & $^7$Li & $^9$Be & $^{12}$C \\ \hline
120 MeV $\pi^+$ & $0.31 \pm  0.03$ & $0.80 \pm 0.05$ & $0.94 \pm 0.04$ & $2.17 
\pm 0.17$ \\
120 MeV $\pi^-$ & & $0.17 \pm 0.02$ & $0.34 \pm 0.02$ & \\
180 MeV $\pi^+$ & $1.05 \pm 0.08$ & $2.43 \pm 0.18$ & $3.15 \pm 0.22$ & \\
180 MeV $\pi^-$ & $0.88 \pm 0.07$ & $0.60 \pm 0.04$ & $0.95 \pm 0.07$ & \\
240 MeV $\pi^+$ & $1.87(1.68) \pm 0.11$ & $4.00(3.75) \pm 0.24$ & 
$4.28(4.03) \pm 0.23$ & $3.73(3.46) \pm 0.25$ \\
240 MeV $\pi^-$ & $1.76(1.57) \pm 0.12$ & $1.36(1.17) \pm 0.09$ & 
$2.22(2.00) \pm 0.10$ & \\
\end{tabular}
\end{ruledtabular}
\end{table}



\begin{table}[tbp]
\caption{Ratios of $^7$Li to $^6$Li total DCX reaction cross sections.}
\label{table:TableV}
\begin{ruledtabular}
\begin{tabular}{c|cccc}
Incident Energy & \multicolumn{2}{c}{\ $R(\pi^+,\pi^-)^a$} 
& \multicolumn{2}{c}{\ $R(\pi^-,\pi^+)^a$} \\ 
(MeV) & Expt. & Scal. Rule & Expt. & Scal. Rule \\
\hline
180 & $2.31 \pm 0.25$ & $1.85$ & $0.68 \pm 0.07$ & $0.69$ \\
240 & $2.14 \pm 0.19$ & $1.85$ & $0.77 \pm 0.08$ & $0.69$
\end{tabular}
\end{ruledtabular}
\footnotetext{$^a$The ratios are $R(\pi^\pm, 
\pi^\mp) \equiv \sigma(\pi^\pm, \pi^\mp)_{^7{\rm Li}} / \sigma(\pi^\pm, 
\pi^\mp)_{^6{\rm Li}}.$}
\end{table}



\begin{table}[tbp]
\caption{Ratios of $(\pi^+, \pi^-)$ to $(\pi^-, \pi^+)$ 
total reaction cross sections for $^7$Li and $^9$Be.}
\label{table:TableVI}
\begin{ruledtabular}
\begin{tabular}{c|cccc}
Incident Energy &  \multicolumn{2}{c}{\ $^7$Li} & \multicolumn{2}{c}{\ $^9$Be} \\ 
(MeV) & Expt. & Scal. Rule & Expt. & Scal. Rule \\
\hline
180 & $4.05 \pm 0.40$ & $2.67$ & $3.32 \pm 0.34$ & $2.08$ \\
240 & $2.94 \pm 0.26$ & $2.67$ & $1.93 \pm 0.14$ & $2.08$
\end{tabular}
\end{ruledtabular}
\end{table}

\pagebreak
\clearpage

\begin{figure}[p]
\begin{center}
\epsfig{file=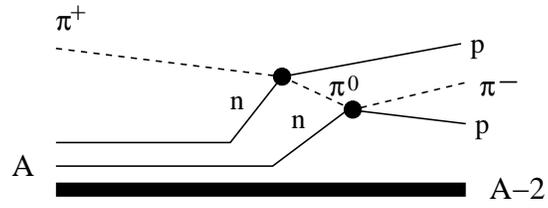,width=0.45\linewidth}
\end{center}
\caption{Schematic diagram of the sequential single charge exchange
mechanism for the $(\pi^+, \pi^-)$ reaction.}
\label{fig1} 
\end{figure}

\begin{figure}[p]
\begin{center}
\epsfig{file=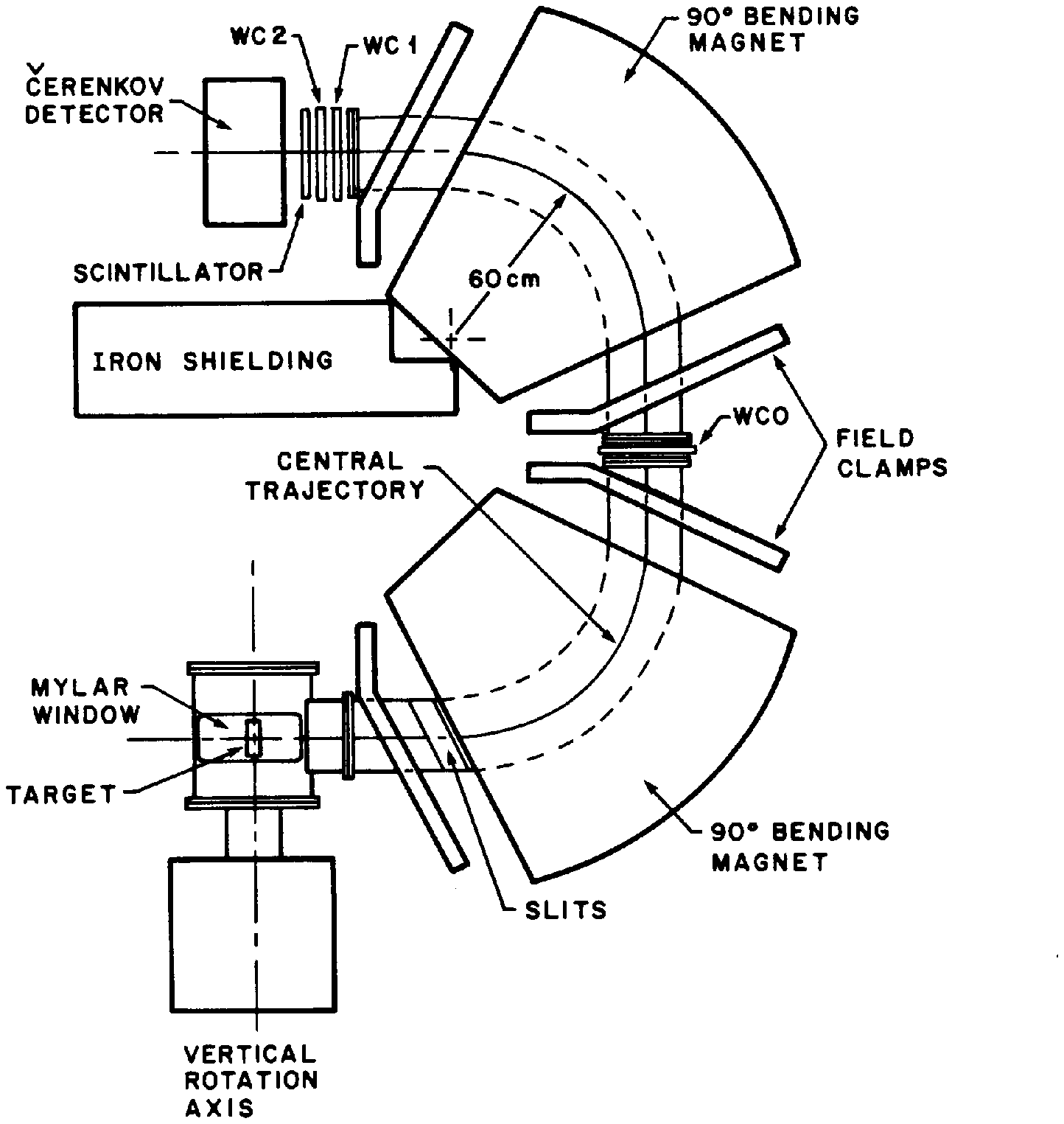,width=\linewidth}
\end{center}
\caption{Drawing of the 180$^\circ$ vertical bend, double focussing
magnetic spectrometer. Pions travel through vacuum from the target, through
two 90$ ^\circ$ dipole magnets, to the focal plane. There is a 2.5
cm break in the vacuum for WC0, the mid-spectrometer wire chamber, which is
used to require that a particle traverses the entire spectrometer.
Particle trajectories are traced back to the focal plane using information
from two wire chambers, WC1 and WC2. The scintillator is used to
distinguish positive 
pions from protons, as well as to provide TOF information. The \v Cerenkov
detector separates pions from electrons and positrons.}
\label{fig2}
\end{figure}

\begin{figure}[p]
\begin{center}
\epsfig{file=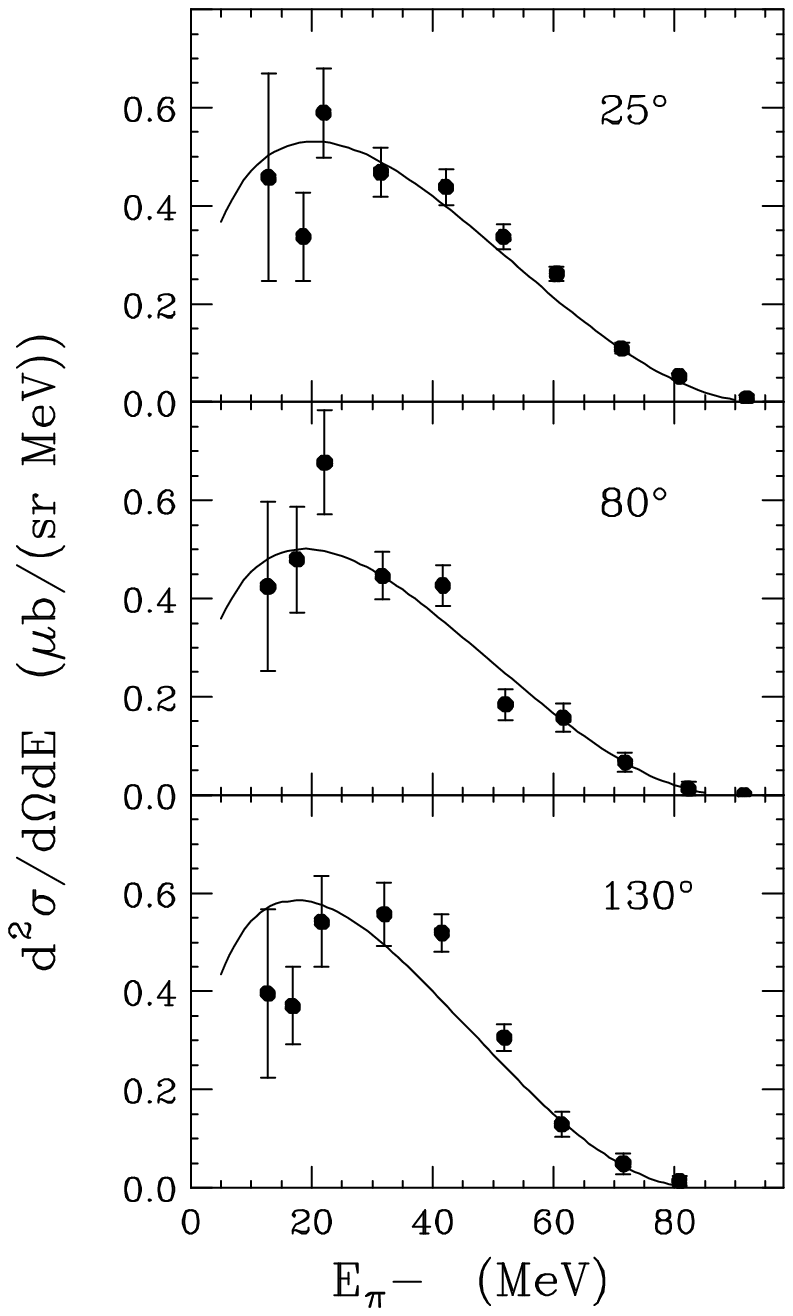,width=0.45\linewidth}
\end{center}
\caption{Doubly differential cross sections for $^6$Li$(\pi^+, \pi^- )$
at incident energy 120 MeV for laboratory angles 
25$^\circ$, 80$^\circ$, and 130$^\circ$. The uncertainties indicated
include the statistical uncertainty and the systematic uncertainties which
depend on the outgoing pion energy.  The solid curves 
represent the distribution of events in four-body phase space.}
\label{fig3} 
\end{figure}

\begin{figure}[p]
\begin{center}
\epsfig{file=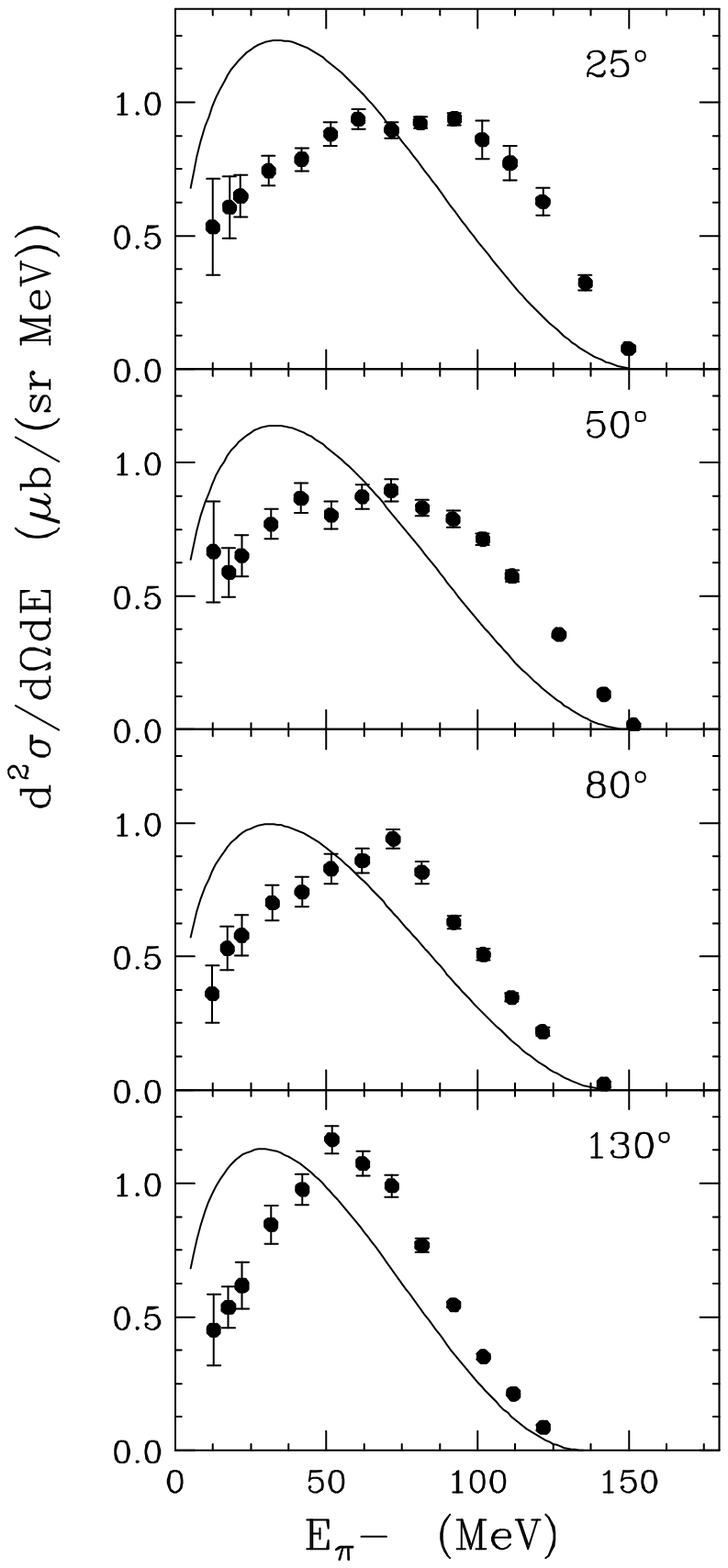,width=0.45\linewidth}
\end{center}
\caption{Doubly differential cross sections for 
$^6$Li$(\pi^+, \pi^- )$ at incident energy 180 MeV for laboratory angles 
25$^\circ$, 50$^\circ$, 80$^\circ$, and 130$^\circ$. The uncertainties 
indicated
include the statistical uncertainty and the systematic uncertainties which
depend on the outgoing pion energy.  The solid curves 
represent the distribution 
of events in four-body phase space.}
\label{fig4}
\end{figure}

\begin{figure}[p]
\begin{center}
\epsfig{file=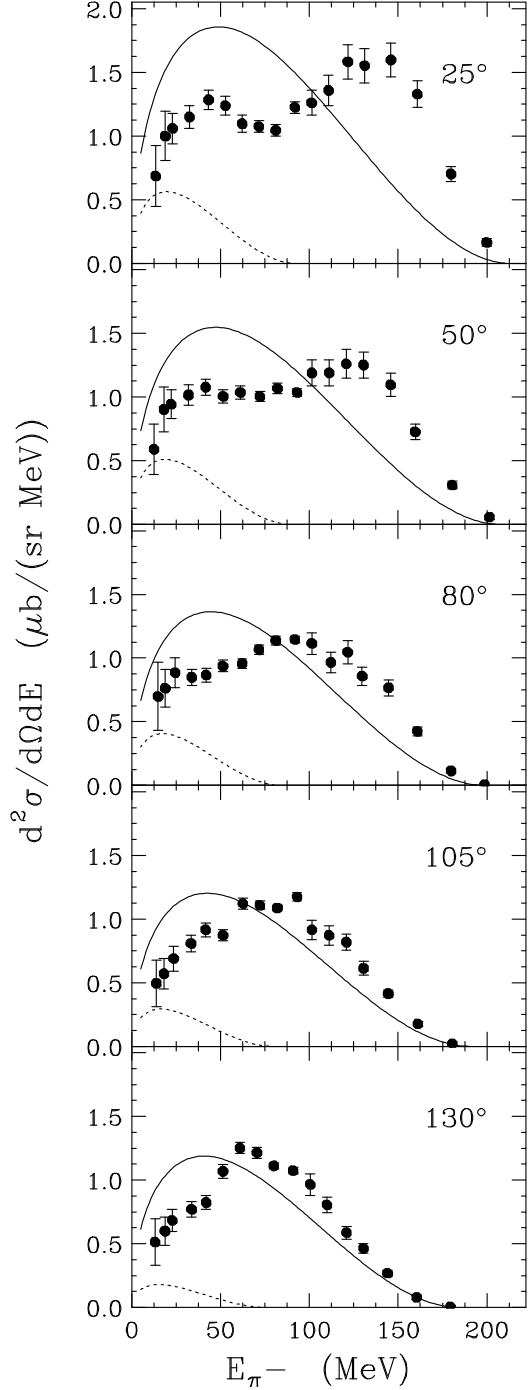,width=0.42\linewidth}
\end{center}
\caption{Doubly differential cross sections for $^6$Li$(\pi^+, \pi^- )$
at incident energy 240 MeV for laboratory angles 
25$^\circ$, 50$^\circ$, 80$^\circ$, 105$^\circ$, and 130$^\circ$. The 
uncertainties indicated include the statistical uncertainty and the 
systematic uncertainties which
depend on the outgoing  pion energy.  The solid curves represent the 
distribution of events in four-body phase space; the dashed curves 
indicate the estimated contributions of PIPP (see text).}
\label{fig5}
\end{figure}

\begin{figure}[p]
\begin{center}
\epsfig{file=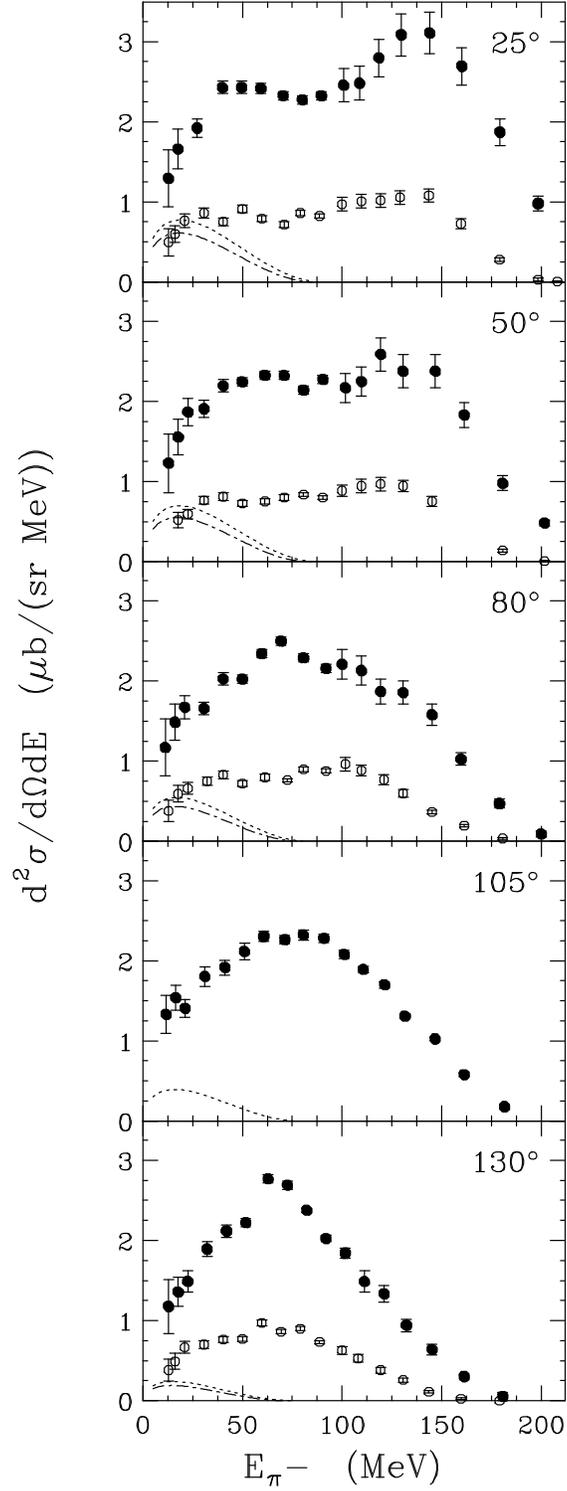,width=0.45\linewidth}
\end{center}
\caption{Doubly differential cross sections for $^7$Li$(\pi^+, \pi^-)$ (solid 
symbols) and $^7$Li$(\pi^-, \pi^+)$ (open symbols) at 240 MeV incident 
energy.  The dotted and dot-dashed curves indicate the estimated 
contributions of PIPP  for $(\pi^+, \pi^-)$ and $(\pi^-, \pi^+)$, 
respectively.} 
\label{fig6} 
\end{figure}

\begin{figure}[p]
\begin{center}
\epsfig{file=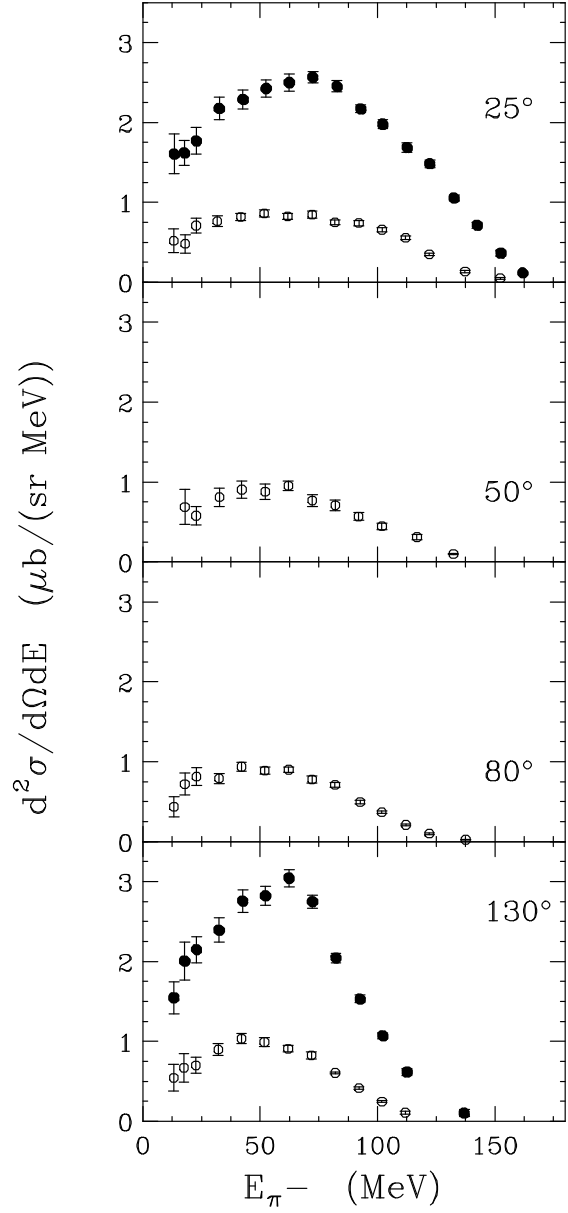,width=0.45\linewidth}
\end{center}
\caption{Doubly differential cross sections for $^9$Be$(\pi^+, \pi^-)$ (solid 
symbols) and $^9$Be$(\pi^-, \pi^+)$ (open symbols) at 180 MeV incident 
energy.} 
\label{fig7}
\end{figure}

\begin{figure}[p]
\begin{center}
\epsfig{file=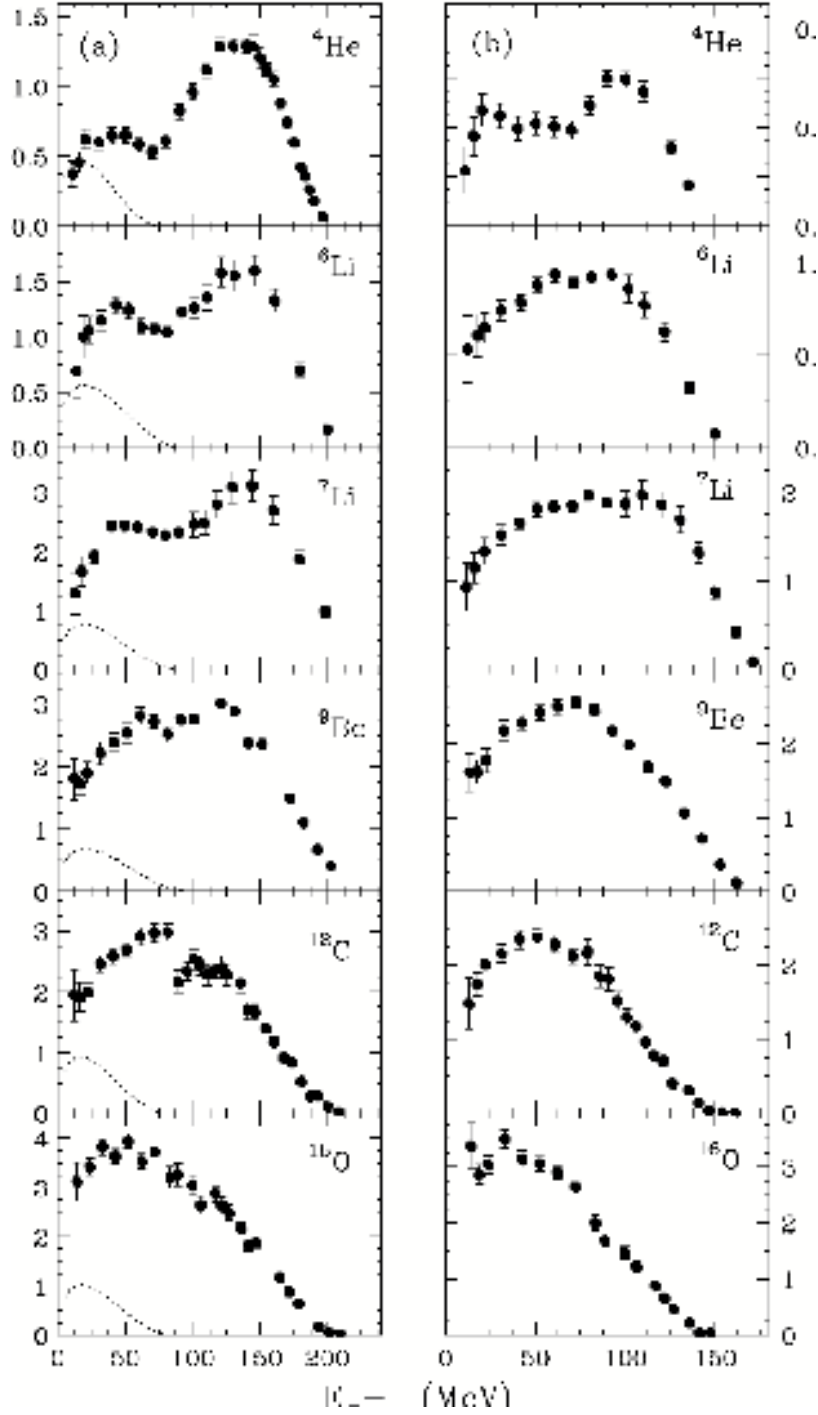,width=0.65\linewidth}
\end{center}
\caption{Doubly differential cross sections at 25$^\circ$ for (a) 240 MeV 
and (b) 180 MeV positive pions incident on $^4$He, $^6$Li, $^7$Li, 
$^9$Be, $^{12}$C, and $^{16}$O. The dotted curves in (a) represent the 
estimated contribution of PIPP (see text).} 
\label{fig8}
\end{figure}

\begin{figure}[p]
\begin{center}
\epsfig{file=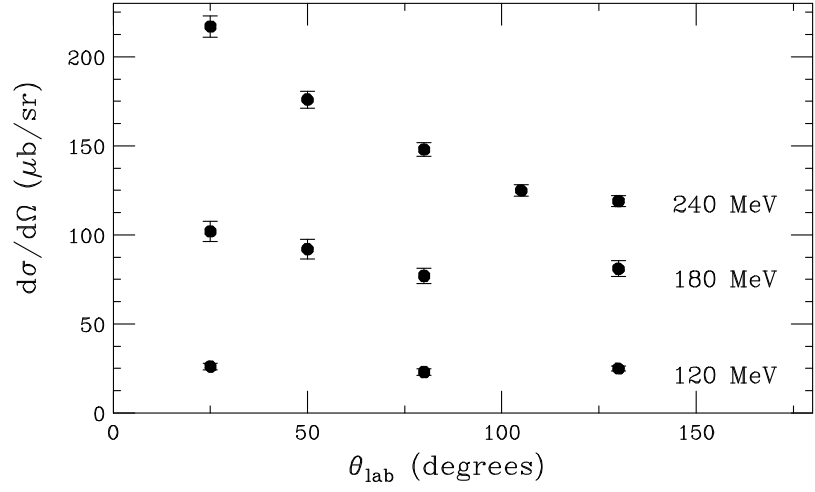,width=0.65\linewidth}
\end{center}
\caption{Angular distributions for $^6$Li$(\pi^+,\pi^-)$ at 120, 180,and 240 MeV. The
uncertainties shown include the statistical uncertainty, the uncertainties arising
from the extrapolation and integration procedure (see text), and the systematic
uncertainties which depend on the outgoing pion energy and angle.}
\label{fig9}
\end{figure}


\begin{figure}[p]
\begin{center}
\epsfig{file=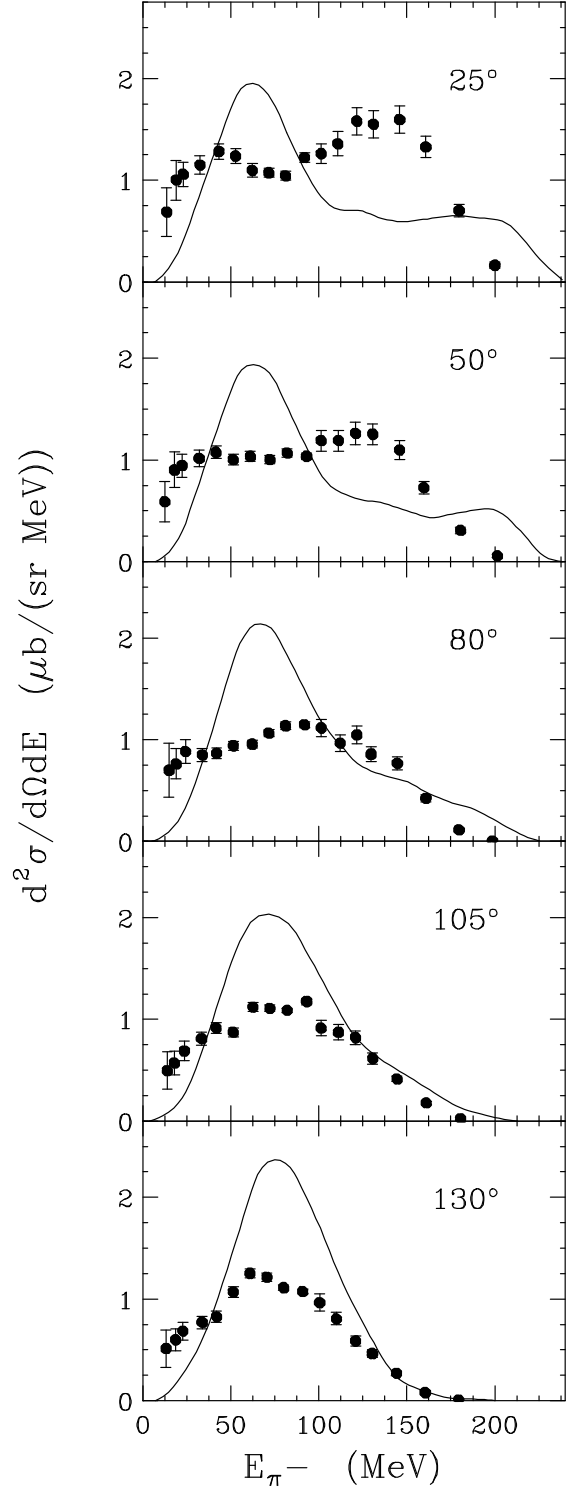,width=0.45\linewidth}
\end{center}
\caption{Comparison of doubly differential cross section for $^6$Li$(\pi^+, 
\pi^-)$ at incident energy 240 MeV with prediction of the intranuclear 
cascade 
calculation of Oset (see text).} 
\label{fig10}
\end{figure}

\begin{figure}[p]
\begin{center}
\epsfig{file=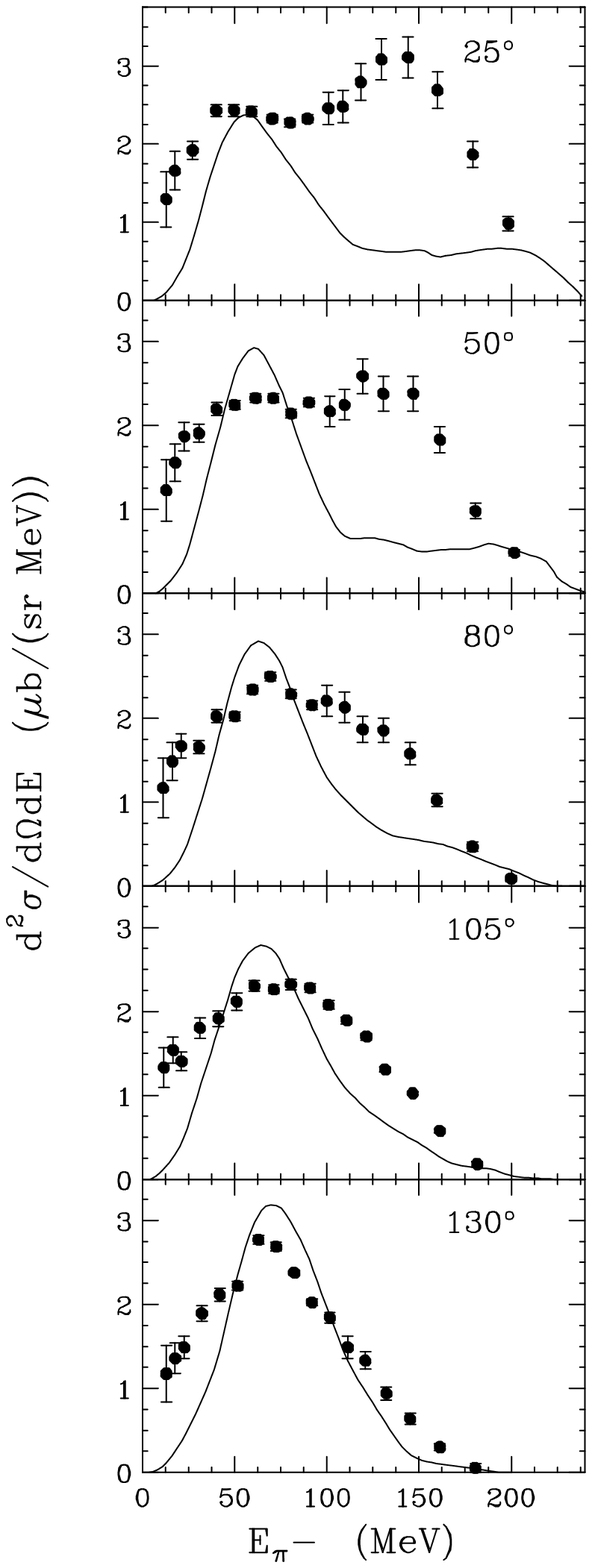,width=0.45\linewidth}
\end{center}
\caption{As Fig.\ 10 for $^7$Li$(\pi^+, \pi^-)$.}
\label{fig11}
\end{figure}

\begin{figure}[p]
\begin{center}
\epsfig{file=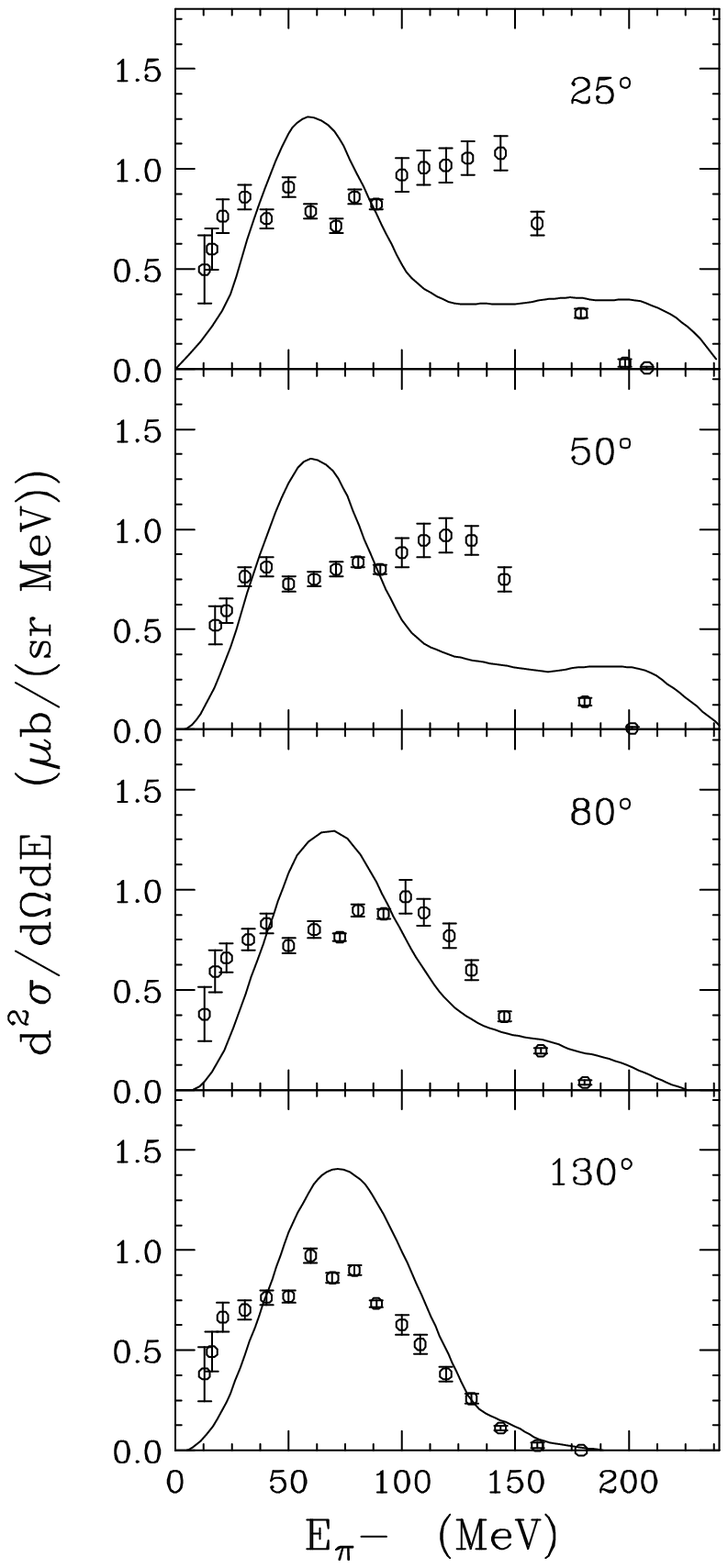,width=0.45\linewidth}
\end{center}
\caption{As Fig.\ 10 for $^7$Li$(\pi^-, \pi^+)$.}
\label{fig12}
\end{figure}

\begin{figure}[p]
\begin{center}
\epsfig{file=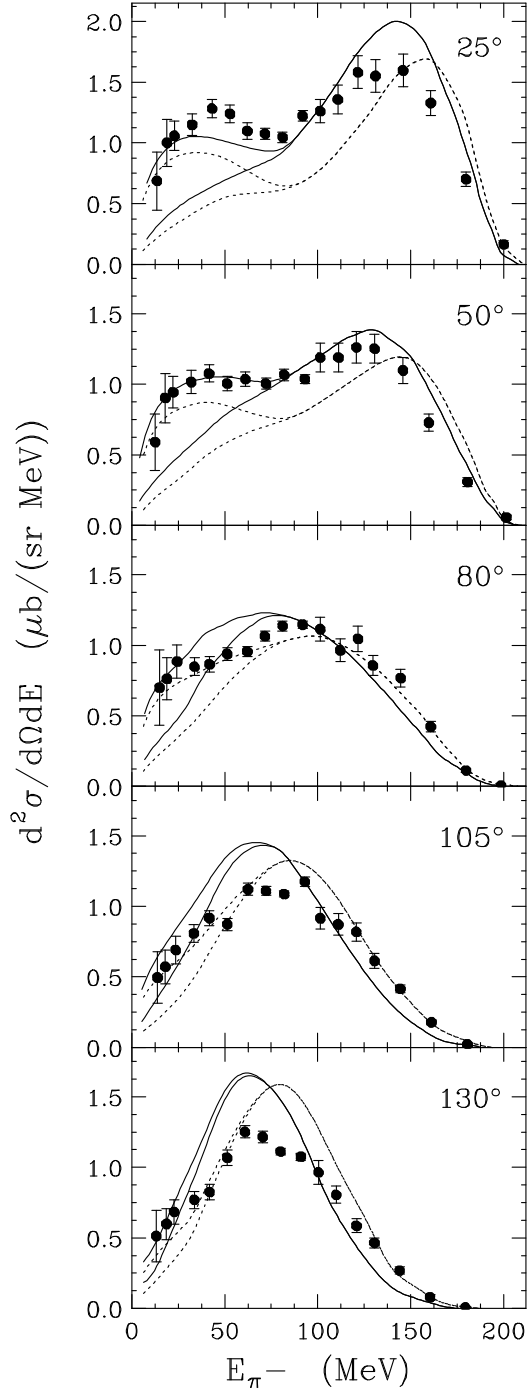,width=0.42\linewidth}
\end{center}
\caption{Comparison of doubly differential cross sections for $^6$Li$(\pi^+, 
\pi^-)$ at incident energy 240 MeV and laboratory angles 25$
^\circ$, 50$^\circ$, 80$^\circ$, 105$^\circ$, and 130$^\circ$ 
with predictions of a non-static sequential 
single charge exchange model (lower branches of curves).  The average 
nuclear potentials are 
$U_1=U_2=-30$
MeV (solid curves), $U_1=U_2=0$ MeV (dotted curves).  The upper branches 
of the curves include PIPP (see text).} 
\label{fig13}
\end{figure}

\begin{figure}[p]
\begin{center}
\epsfig{file=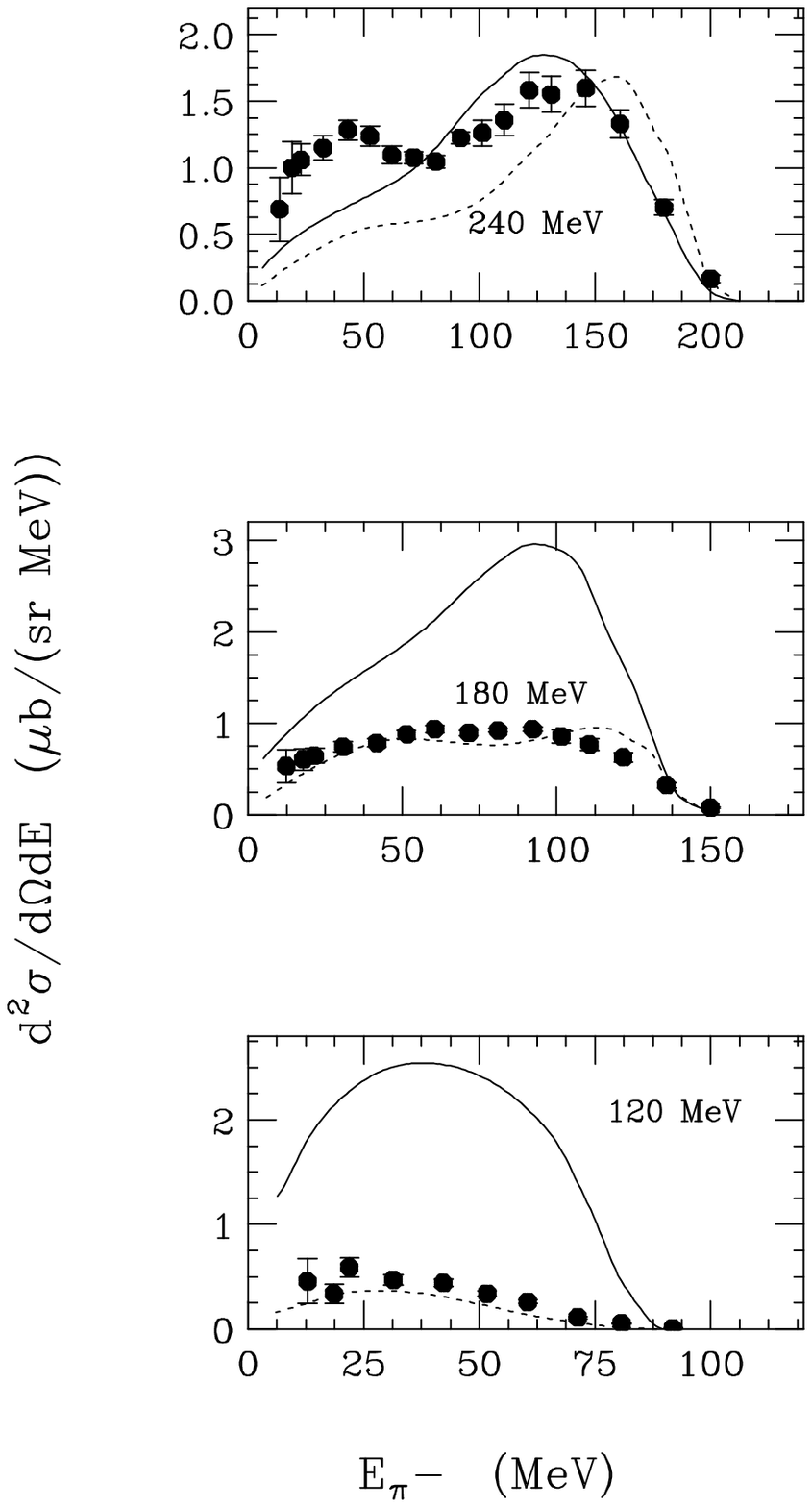,width=0.45\linewidth}
\end{center}
\caption{Comparison of doubly differential cross section for $^6$Li$(\pi^+, \pi^-)$ at 
25$^\circ$ for 
incident energies 120, 180, and 240 MeV with prediction of a non-static 
sequential single charge exchange model The curves are as in Fig. 13.} 
\label{fig14}
\end{figure}


\begin{thebibliography}{99}

\bibitem{grametal} P.~A.~M. Gram, S.~A. Wood, E.~R. Kinney, S. H\o
ibr\aa
ten, P. Mansky, J.~L. Matthews, T. Soos, G.~A. Rebka, Jr., and D.~A.
Roberts, Phys. Rev. Lett. {\bf 62}, 1837 (1989).

\bibitem{mischke} R.~E. Mischke, A. Blomberg, P.~A.~M. Gram, J. Jansen, J. 
Zichy, J. Bolger, E. Boschitz, C.~H.~Q. Ingram, and G. Pr{\" o}bstle,
Phys.  Rev. Lett. {\bf 44}, 1197 (1980). 

\bibitem{wood}  S.~A. Wood, J.~L.~Matthews, E.~R. Kinney, P.~A.~M. Gram,
G.~A. Rebka, Jr., and D.~A. Roberts, Phys. Rev. C {\bf 46}, 1903 (1992).

\bibitem{yuly} M.~Yuly, W.~Fong, E.~R. Kinney, C.~J.~Maher,
J.~L.~Matthews, T.~Soos, J.~Vail, M.~Y.~Wang, S.~A.~Wood, P.~A.~M.~Gram, 
G.~A.~Rebka, Jr. and D.~A.~Roberts, Phys.~Rev.~C {\bf 55}, 1848 (1997).

\bibitem{kprl} E.~R. Kinney, J.~L. Matthews, P.~A.~M. Gram, D.~W. 
MacArthur, E. Piasetzky, G.~A. Rebka, Jr., and D.~A. Roberts, Phys. Rev.
Lett. {\bf 57}, 3152 (1986). 

\bibitem{kth}  E.~R. Kinney, Los Alamos National Laboratory Report No.
LA-11417-T, 1988.

\bibitem{kinney}  E.~R. Kinney, J.~L. Matthews, P.~A.~M. Gram, D.~W.
MacArthur, E. Piasetzky, G.~A. Rebka, Jr., and D.~A. Roberts, Phys.~Rev.~C {\bf 72}, 
044608 (2005).

\bibitem{fongth} W. Fong, Ph.D. Thesis, Massachusetts Institute of
Technology, 1994.

\bibitem{fong} W. Fong, J.~L. Matthews, M.~L. Dowell, E.~R. Kinney, S.~A.
Wood, P.~A.~M. Gram, G.~A. Rebka, Jr., and D.~A. Roberts, Few-Body Systems 
Suppl. {\bf 9}, 187 (1995).

\bibitem{gil64} L. Gilly, M. Jean, R. Meunier, M. Spighel, J.~P. 
Stroot, P. Duteil, and A. Rode, Phys. Lett. {\bf 11}, 244 (1964).

\bibitem{gil65} L. Gilly, M. Jean, R. Meunier, M. Spighel, J.~P. 
Stroot, and P. Duteil, Phys. Lett. {\bf 19}, 335 (1965).

\bibitem{bat66} Yu.~A. Batusov, S.~A. Bunyatov, V.~M. Sidorov, and 
V.~A. Yarba, Sov. J. Nucl. Phys. {\bf 3}, 223 (1966).

\bibitem{mas71} J.~P. Massu{\' e}, Y. Sakamoto, Yu.~A. Batusov, and P. 
C{\" u}er, Nucl. Phys. {\bf B29}, 
515 (1971).

\bibitem{evs81} V.~S. Evseev, V.~S. Kurbatov, V.~M. Sidorov, V.~B. 
Belyaev, J. Wrzecionko, M. Daum, R. Frosch, J. McCulloch, and E. 
Steiner, Nucl. Phys. {\bf A352}, 379 (1981).

\bibitem{abramov} B. M. Abramov, L. Alvarez-Ruso, Yu. A. Borodin, S. A. 
Bulychjov, M. J. Vicente Vacas, I. A. Dukhovskoy, A. P. Krutenkova, V. V. 
Kulikov, M. A. Matsyuk, and E. N. Turdakina, Yadernaya Fizika {\bf 68}, 
1336 (2005) [Physics of Atomic Nuclei {\bf 68}, 1283 (2005)].

\bibitem{walter}  J. B. Walter and G. A. Rebka, Jr., 
Los Alamos National Laboratory Technical Report No. LA-7731-MS (1979).

\bibitem{wth}  S.~A. Wood, Los Alamos National Laboratory Report No.
LA-9932-T, 1983.

\bibitem{ashery} D.~Ashery, I.~Navon, G.~Azuelos, H.~K.~Walter, 
H.~J.~Pfeiffer, and F.~W. Schlep{\"utz, }Phys. Rev. C {\bf 23}, 2173 (1981).

\bibitem{bjork} C.~W.~Bjork {\it et al.}, Phys. Rev. Lett. {\bf 44}, 62 (1980).

\bibitem{piasetzky} E.~Piasetzky, P.~A.~M.~Gram, D.~W.~MacArthur, G.~A.~Rebka, 
Jr., C.~A.~Bordner, S.~Hoibraten, E.~R.~Kinney, J.~L.~Matthews, S.~A.~Wood, and 
J.~Lichtenstadt, Phys. Rev. Lett. {\bf 53}, 540 (1984); J.~Lichtenstadt, 
D.~Ashery, S.~A.~Wood, E.~Piasetzky, P.~A.~M.~Gram, D.~W.~MacArthur, 
R.~S.~Bhalerao, L.~C.~Liu, G.~A.~Rebka, Jr. and D.~Roberts, Phys. Rev. C {\bf 
33}, 655 (1986).

\bibitem{chaos} F. Bonutti {\it et al.}, Nucl. Phys. {\bf A677}, 213 
(2000).

\bibitem{rahav} A.~Rahav {\it et al.}, Phys. Rev. Lett {\bf 66}, 1279 (1991).

\bibitem{grion} N.~Grion {\it et al.}, Nucl. Phys. {\bf A492}, 509 (1989).

\bibitem{ag} M.~Alqadi and W.~R.~Gibbs, Phys. Rev. C {\bf 65}, 044609 
(2002).

\bibitem{buss} O. Buss, L. Alvarez-Ruso, A. B. Larionov, and U. Mosel, 
Phys. Rev. C {\bf 74}, 044610 (2006).

\bibitem{oset} E. Oset, L.~L. Salcedo, and D. Strottman, Phys. Lett. B
{\bf 165}, 13 (1985).

\bibitem{salcedo} L.~L. Salcedo, E. Oset, M.~J. Vincente-Vacas, and C. 
Garc\'{i}a-Recio Nucl. Phys. {\bf A484}, 557 (1988).

\bibitem{vicente} M. J. Vicente, E. Oset, L. L. Salcedo, and C. Garc\'{i}a-Recio,
Phys. Rev. C {\bf 39}, 209 (1989). 

\bibitem{vo} M. Vicente-Vacas and E. Oset, in {\em
Second LAMPF International Workshop on Pion-Nucleus Double Charge Exchange}, Los
Alamos, 1989, edited by W.~R. Gibbs and M.~J. Leitch (World Scientific, Singapore,
1990), p. 120.

\bibitem{vl} J. van Loon, Master's Thesis, Free University, Amsterdam,
1985. 

\bibitem{thies}  M. Thies, Ann. Phys. (N.Y.) {\bf 123}, 411 (1979).


\end{thebibliography}
\end{document}